\documentclass[aps,prd,preprint,amsmath,amssymb,graphicx,longbibliography]{revtex4-1}
\usepackage{amsmath}
\usepackage[caption=false]{subfig}
\usepackage{bbm}
\usepackage[utf8x]{inputenc}
\usepackage[textsize=tiny]{todonotes}

\usepackage{appendix}
\usepackage{xcolor}
\usepackage{float}
\usepackage{braket}
\usepackage{slashed}
\usepackage{graphicx}
\definecolor{darkred}{rgb}{0.4,0.0,0.0}
\definecolor{darkgreen}{rgb}{0.0,0.3,0.0}
\definecolor{darkblue}{rgb}{0.0,0.0,0.7}
\usepackage[bookmarks,linktocpage,colorlinks,
linkcolor = darkred,
urlcolor  = darkgreen,
citecolor = darkblue]{hyperref}
\def\beq{\begin{equation}}
  \def\enq{\end{equation}}
\usepackage{lineno}
\usepackage{comment}
\definecolor{winered}{rgb}{0.8,0,0}
\definecolor{darkb}{rgb}{0,0,0.8}

\usepackage[english]{babel}

\begin{document}

\title{Tensor networks for High Energy Physics: contribution to Snowmass 2021} 
\author{Yannick Meurice$^{1}$}
\author{James C. Osborn$^2$}
\author{Ryo Sakai$^{3}$}
\author{Judah Unmuth-Yockey$^{4}$}
\author{Simon Catterall$^{3}$}
\author{Rolando D. Somma$^{5}$}
\affiliation{$^1$ Department of Physics and Astronomy, The University of Iowa, Iowa City, IA 52242 USA }
\affiliation{$^2$ Computational Science Division, Argonne National Laboratory, Argonne, IL 60439}
\affiliation{$^3$ Department of Physics, Syracuse University, Syracuse, NY 13244 USA}
\affiliation{$^4$ Fermilab, Batavia, IL 60510}
\affiliation{$^5$ Los Alamos National Laboratory, Los Alamos, NM 87545}
\def\beq{\begin{equation}}
  \def\enq{\end{equation}}

\date{\today}

\begin{abstract}
Tensor network methods are becoming increasingly important for high-energy physics, 
condensed matter physics
and quantum information science (QIS). We 
discuss the impact of tensor network methods on lattice field theory, quantum gravity and  
QIS in the context of High Energy Physics (HEP). These tools 
will target calculations for strongly interacting systems that are made difficult 
by sign problems when conventional Monte Carlo and other importance sampling methods are used.
Further development of
methods and software will be needed to make a significant impact in HEP. 
We discuss the roadmap 
to perform quantum chromodynamics (QCD) related calculations in the coming
years. The research is labor
intensive and requires state of the art 
computational science and computer science input
for its development and validation.
We briefly discuss the overlap with other science domains and industry.
\end{abstract}

\maketitle
\tableofcontents{}

\newpage
\centerline{\bf Executive summary for “CompF6: Quantum computing”}

Tensor network methods are playing an increasingly 
important role in several branches of physics and in quantum information science.
Tensor networks provide a compact representation of entangled quantum systems and quantum states.
They can be used to reformulate physical theories on a lattice and map them onto other quantum
systems---such as quantum computers or quantum simulators. These physical theories can then either be studied classically using the original tensor network formulation, or in the mapped representation using quantum hardware.

In the original formulation, they present an increasingly 
attractive alternative to traditional Monte Carlo methods (used in lattice QCD)
which are currently the only computational methods able to
provide first-principle calculations for a large range of HEP problems,
but which also consume large computational resources. 
In numerical methods based on coarse-graining, 
tensor network methods do not rely on statistical sampling, potentially providing solutions
to problems plagued by \emph{sign problems} that render Monte Carlo methods ineffective.
In some fields, such as condensed matter physics, their use is fairly common.
However, for problems of interest to HEP, 
tensor network methods are still very resource intensive, and are not currently feasible
for complex theories in four space-time dimensions, such as QCD.
Further development of tensor network methods and software is needed for them to make
a significant impact in HEP like Monte Carlo.

In addition to providing an alternative to Monte Carlo methods in HEP, tensor networks are also useful in connecting HEP to QIS.   They can provide a convenient representation for mapping lattice theories to quantum hardware (either in terms of digital quantum circuits or
analog quantum simulators),
and can also be used to  perform classical simulations of quantum circuits.
This is helpful for developing and testing quantum computing algorithms, and is an important
area of research for determining the tipping point for quantum computers to provide a
computational advantage over classical ones.
They also assist in isolating ``building blocks'' 
that have limited complexity and can be optimally approximated for noisy intermediate-scale quantum (NISQ) hardware. 

Here we summarize some of the key problems in HEP that could be addressed with tensor network
methods, the current status of applications to HEP, 
and the developments in methods, software, and personnel needed to make progress
in these areas.

\newpage
\centerline{\bf Executive summary for “TheoryF10: Quantum Information Science"}

Tensor network methods are playing an 
increasingly important role in several branches of physics and in quantum information science.
In the context of HEP, tensor networks have appeared as ways to reformulate lattice gauge theory models to obtain a fully discrete formulation that is suitable 
for quantum computation and coarse graining methods. Tensor networks 
also provide tools to understand entanglement in conformal field theories and their connection to gravity. 

In the context of lattice gauge theory, tensors can be seen as the 
translationally invariant, local {\it building blocks} of exact 
discretizations of the path integral. 
They encode both the local and global symmetries of the original model. It is easy to design approximations (truncations) that preserve these symmetries and to design simplified 
models that should have the same correct universal continuum limit as the 
original model. Developing these building blocks and optimizing the approximations for NISQ machines and classical computers are important tasks for the near-term future. 
Tensor networks can also be used to perform 
classical simulations of quantum circuits.
This is useful for developing and testing quantum computing algorithms and quantum computational advantage.

There is an emerging international and interdisciplinary 
community developing new methods in this area, which includes an increasing number of researchers from 
the lattice gauge theory community. There is a clear road map to do QCD related calculations in the coming years. This effort is at the interface of quantum computing and classical high performance
computing (HPC) and
requires state of the art HPC for its development and validation.

\newpage
\centerline{\bf Executive summary for “TheoryF5: Lattice Gauge Theory"}

Tensor Lattice Field Theory (TLFT) is a 
new approach for understanding the non-perturbative structure of lattice theories including
QCD.
Starting from the conventional 
Lagrangian/path-integral formulation, it is possible to reformulate the models using 
tensors which are translationally invariant, local, and 
can be considered as {\it building blocks} of exact 
discretizations of the path integral. They encode the local and global symmetries of the original model. They provide a more general space of theories than the 
traditional Lagrangian and Hamiltonian densities. 
Most models studied by lattice gauge theorists can be reformulated 
in the language of TLFT. 
Truncations need to be used for 
practical purposes but can chosen to preserve symmetries in a generic way. This provides a broad class of models with the same universal continuum limit as the original model.

TLFT smoothly connects the Lagrangian and Hamiltonian approaches and is very useful in the 
context of quantum computing. Moreover, TLFT approaches to the path
integral can also be combined
with deterministic coarse-graining methods. 
This is very efficient for a broad range of couplings and also allows the study
of complex actions. 
The truncations performed during coarse-graining are the only approximations used and,
if they can be controlled, exponentially large volumes can be reached. 
This is potentially interesting in the study of models beyond the standard model with slowly running coupling constants.

Using TLFT 
for models directly relevant for HEP is a feasible goal in the next decade. 
Developing software and running codes in TLFT is labor intensive and requires access to high
performance computing.

\newpage
\section{Introduction}
The use of tensor network methods to tackle quantum field theory problems relevant 
for HEP is a recent and rapidly developing new area of research.
This is a very interdisciplinary area and workshops 
involving several communities have taken place recently.
For instance, a recent workshop at the Institute for Nuclear Theory \cite{int21-1c} had more than 80 participants in High Energy, Nuclear and Condensed Matter Physics.
Before discussing the science goals, the importance of the approach for quantum computation,
recent progress, and resources needed, we first give a brief overview of the context in which tensor network methods appear, and their relations with other areas of research.

An important disclaimer: this white paper does not provide exhaustive lists of 
references but rather selected references that are most familiar to tensor 
practitioners in the US lattice community. Exhaustive lists of references can be found in recent reviews \cite{Orus_2019,Banuls:2019bmf,Banuls:2019rao,rmp,RevModPhys.93.045003}. Additional information can also be found in some of the 2020 LOI's \cite{yloi,jloi,nloi}. 
\begin{itemize}
\item
{\bf Context} \\
Tensor network methods are playing an increasingly important role in 
many branches of physics and data analysis. In the context of HEP and 
more specifically lattice gauge theory, 
tensor network methods appear at the interface of quantum and classical computing and 
target questions where sign problems are present or large volumes are needed. They can also
play a role in understanding conformal and holographic theories.
For recent reviews focused on the use of tensors in lattice gauge theory see \cite{Banuls:2018jag,Banuls:2019rao,rmp}. 
\item
{\bf Terminology}\\ 
The term ``tensor" appears in many 
other contexts. For instance,  
energy-momentum tensors or  Kalb-Ramond tensor fields are often used in HEP. 
In contrast, ``tensor networks" 
refers to objects with multiple indices that can be assembled together 
to represent the states of a 
Hilbert space or transfer matrix, compute expectation values of operators, or express partition functions.
Typically, the indices refer to quantum numbers or to  
a list of states in a localized Hilbert space.
\item
{\bf Relationship with quantum computing} \\
The reformulations of lattice field theory models using 
finite sets of tensor indices provide Hilbert spaces suitable for quantum computing (see Section \ref{sec:relation}). The space-time tensor networks can be adapted to build quantum circuits. Additionally, certain
quantum states that can be represented using a tensor network, e.g. matrix product states (MPS), 
can be prepared efficiently on a quantum computer. This allows one to address more general problems, e.g., quantum simulation, which are still inefficient for classical computers. 

\item
{\bf Condensed matter origins} \\
Many of the basic quantum information ideas behind the use of tensor networks 
in lattice field theory were originally developed in the context of condensed matter. 
White emphasized the need to keep track of the entanglement among 
coarse-grained blocks while performing real space renormalization group
transformations and invented the density matrix renormalization group (DMRG) method to handle
this \cite{PhysRevLett.69.2863}. DMRG is typically the most accurate method for computing ground states of one-dimensional lattice Hamiltonians and is related to 
tensor networks called matrix product states (MPS). Other tensors networks, for instance, called projected entangled pair states (PEPS) and the multi-scale entanglement renormalization ansatz (MERA), play an important role in dealing with other types of models.
The quantum information theory 
community has developed a clear understanding of the 
convergence (or the lack thereof) of such algorithms in terms of area laws 
associated with entanglement entropy.  For recent reviews see \cite{hv,ran}. 
\item
{\bf Tensors for HEP and nuclear physics}\\ 
There has been a significant interdisciplinary effort 
starting around 2010, and highlighted in conferences and 
workshops at the Aspen Center for Physics, KITP, the INT and KITPC, to apply some of
methods developed in the condensed matter community to problems in HEP and nuclear physics (NP).
The annual lattice conferences have helped foster interactions among the communities involved 
 \cite{mcbpos2013,ympos2013,kvapos2014,yspos2014,mcbpos2014,juypos2014,ympos2014}. 
 The number of contributions has grown steadily with the years. 
 Tensor reformulations of lattice gauge theory models starting from the standard Lagrangian path-integral formalism connect 
 smoothly to Hamiltonian approaches developed in condensed matter physics. 
\item
{\bf Tensors as building blocks of computations}\\
The tensors are local objects containing  
all the universal information about the model in question such as its dimension and symmetries.  
Most lattice models have a tensor reformulation
and there is a clear plan to follow the road 
map that has been successful for QCD (the ``Kogut sequence", see below).
Tensors can be used to deal with sign problems, strongly interacting systems and 
conformal field theory (but efficient numerical methods remain to 
be fully developed in 2+1 and 3+1 dimensions). They provide 
new tools for the lattice field theory and AdS/CFT communities.
Quantum circuits can also be viewed as tensor networks, providing alternative methods
for constructing, transforming and evaluating quantum computations \cite{doi:10.1137/050644756,PhysRevLett.91.147902,biamonte2017tensor}.

\item
{\bf The road map}\\
The lattice gauge theory community 
worked its way over many years to being able to simulate 
full QCD with dynamical fermions by starting with simpler models in lower dimensions. This approximate sequence of models is sometimes called 
the ``Kogut sequence" or the ``Kogut ladder" after review articles \cite{kogut79,kogut83}.
A similar road map is being followed with the tensor reformulations discussed here \cite{rmp}. 

\end{itemize}

\section{Science goals for HEP}
\begin{itemize}
\item
{\bf Quantitatively reliable coarse-graining for strong interactions}\\
Quarks and gluons play an important role for 
problems with typical energies ranging from MeV to TeV. Hypothetical models beyond the standard 
model often involve effective coupling constants that run very slowly with energy
scale. To tackle such problems
it is crucial to develop reliable methods for performing coarse-graining and
renormalization. Tensor renormalization group (TRG)  methods have the potential to meet these expectations because they allow a clean 
partition of the degrees of freedom at different scales. Indeed such methods
have been used in simple models to obtain critical exponents to very high precision.
However, the
truncation methods that must be employed still need optimization for use
in higher dimensions with both fermions and gauge fields. 
\item
{\bf Dealing with sign problems at finite density}\\ 
A certain number of problems in astrophysics and nuclear physics 
require calculations involving quarks and gluons at finite density. 
These calculations are plagued by sign problems 
which prevent the use of conventional importance sampling. 
TRG methods rely on the singular value decomposition (SVD) and 
are typically insensitive to sign problems. 
In 1+1 dimensions, they are generically very 
efficient in dealing with problems involving finite density or 
complex couplings where the Euclidean Boltzmann weights are 
not real and positive. Developing efficient methods in 
higher dimensions is an important goal for the tensor community. 
\item
{\bf Real-time evolution of strongly interacting particles}\\
The interpretation of hadron collider data 
relies extensively on event generation algorithms such as Pythia \cite{pythia}. 
These algorithms incorporate results from 
perturbative QCD that are reliable at short distances and use empirical models to 
describe the formation of hadrons at larger distances. Replacing these empirical models 
by ab-initio calculations based on lattice QCD is one of the 
major motivations for quantum computing in HEP. This is discussed in a separate white paper
\cite{simulwp}. 
Tensor methods can also in principle be used to handle 
real-time evolution and out-of equilibrium situations. This is a well-developed research area in condensed matter and we hope that this can also become the case in HEP. 

\end{itemize}

\section{Relation with Quantum Computing and QIS}
\label{sec:relation}
\begin{itemize}
\item
{\bf Discretization of path-integrals}\\ 
For qubit-based quantum computing a complete discretization of both field
space and spacetime is necessary. 
Tensor network methods starting from a lattice path-integral 
formulation provide a general way to reach this goal. 
\item
{\bf Character expansions}\\
For compact groups, character expansions (e.g., Fourier expansion) 
provide a natural way to discretize the
path integral of models with continuous variables (e.g., the gluon fields in QCD). They were extensively developed in the context of strong coupling expansions in the early days
of lattice QCD but they also connect with modern worm algorithms and can be adapted to
work well even at weak coupling.
\item
{\bf Symmetry compatible truncations}\\ 
For practical purposes, truncations of the character expansions are necessary. 
Fortunately, this can be done in a way compatible with symmetries \cite{rmp,ymdis,ymex}.
\item
{\bf Locality and effective theories}\\ The (quasi) locality of microscopic
interactions is a crucial ingredient of Lloyd's quantum supremacy argument \cite{lloyd1996universal}. It is present in tensor network reformulations and their effective theories obtained by coarse graining.
\item
{\bf Simulation of the building blocks of quantum algorithms}\\
Quantum algorithms can be interpreted as tensor networks, where each tensor corresponds to a quantum gate. From this point of view, we can isolate building blocks of the
quantum algorithm and use tensor network methods 
to perform the tensor contractions,
providing a useful technique to simulate quantum algorithms on classical computers.  
\item
{\bf State preparation}\\
Tensor networks with some local structure 
allow one to produce state-preparation quantum circuits of small size or depth. 
This is beneficial for quantum simulation. It is also beneficial to study/verify new 
physical theories and for observing convergence by changing the bond dimension of the tensor. 
\item
{\bf Quantum state tomography}\\
Tensor networks can also be used in the context of quantum state tomography \cite{cramer2010}. By imposing a tensor network structure (e.g., a PEPS), 
there is a way
perform efficient quantum state tomography, which requires exponential resources in general. Efficient quantum state tomography is 
important for many reasons, including the
verification of quantum computing devices and experiments.
\item
{\bf Relation with other approaches}\\
The study of the transfer matrix derived from the tensor 
formalism leads to finite algebras \cite{ymex} that can be compared 
with quantum link constructions \cite{orland1990,brower1997}. Related methods 
\cite{buser2020,bbk} have been used in the context of quantum computing.
These related formulations should be compared in the continuum limit.
For instance, it has been showed recently 
that truncations of the
Abelian Higgs model in the charge representation 
leads to a phase transition associated with an enhanced symmetry.

\end{itemize}

\section{Selected recent progress}

\begin{itemize}
\item {\bf
Spin and gauge models with Abelian symmetries}\\
It was found out early \cite{exact2} that the character expansions for Abelian 
groups leads to simple factorization. The integration of the fields result 
in Kronecker deltas which represent the symmetry (modulo $n$ for $Z_n$ versions). 
The identities representing the symmetry 
depend only on these selection rules and not on 
the specific values taken by the tensor elements \cite{ymex,ymdis}.
Consequently, truncations preserve symmetries exactly and it is possible to 
attempt to reach the continuum limit with highly simplified 
microscopic formulations. In addition, for gauge theories, one can integrate completely the gauge fields without gauge fixing and the procedure is manifestly gauge invariant. 

\item {\bf Spin and gauge models with non-Abelian symmetries}\\
The mass-gap of the O(3) nonlinear sigma model in 2D was studied 
in Ref.~\cite{Unmuth-Yockey:2015Gg} by fitting the two-point correlation function.  
In Refs.~\cite{PhysRevD.93.114503} the CP(1) model in 2D with and without a $\theta$-term was
analyzed using the higher order tensor renormalization group, and the loop-tensor network renormalization algorithms.  
Reference~\cite{PhysRevD.99.114507} considered the non-Abelian Higgs model in 2D.  This model is confining across the entire phase diagram, and the string tension was extracted from Polyakov loop correlators.
Extracting spectra and binding energies from correlations 
functions is an important step on the path towards analyzing QCD.  
Understanding the systematics of the method in the presence of a $\theta$-term --
which has a sign problem  -- for a theory 
which is asymptotically free will help guide 
the method in higher dimensions with other non-Abelian groups.
\item {\bf Scalar theories} \\
The complex scalar $\phi^4$ theory with finite chemical potential, a typical model with the sign problem, is analyzed in 2D and 4D~\cite{Kadoh:2019ube,Akiyama:2020ntf}.
While the presence of a sign problem is a strong motivation to use tensor network methods, 
a notable accuracy of the critical coupling constant compared to other schemes 
including Monte Carlo simulations is reported in the 2D real $\phi^{4}$ theory~\cite{Kadoh:2018tis,Delcamp:2020hzo} following the pioneering work by Shimizu~\cite{Shimizu:2012wfa}.
To discretize the field space a Taylor expansion of the hopping factors 
combined with Gaussian quadratures is used.
\item {\bf Fermions} \\
Application of the Grassmann tensor renormalization group~\cite{Gu:2010yh,Gu:2013gba} to 
relativistic fermion models started from a series of investigation on the 2D 
Schwinger model with Wilson fermions~\cite{Shimizu:2014uva,Shimizu:2014fsa,Shimizu:2017onf}.
A key feature of these fermion models is the presence of Grassmann variables that
decorate the tensor networks.
In some special cases that include the 2D Schwinger model with staggered fermions, it is shown that one can construct a tensor network representation without Grassmann variables~\cite{Butt:2019uul}.
A coarse-graining algorithm for Grassmann variables in higher dimensions that goes well with the higher order TRG (HOTRG)~\cite{2012PhRvB..86d5139X} and its approximation schemes~\cite{Adachi:2019paf,Kadoh:2019kqk} was proposed in~\cite{Sakai:2017jwp}.
\item {\bf Supersymmetric models} \\
The 2D $\mathcal{N}=1$ Wess--Zumino model, the simplest supersymmetric model with a severe sign problem, is analyzed using the Grassmann tensor renormalization group~\cite{Kadoh:2018hqq}.
In this work, a non-interacting case that is exactly solvable is employed as a numerical test.
More complicated models including the interacting case of the same model and the $\mathcal{N}=(2,2)$ Wess--Zumino model would be a possible and interesting direction.
\item {\bf 
Quantum gravity} \\
Tensor networks can be useful in the study of quantum gravity as well.  One of the directions where tensor networks have appeared is in the ``spin-foam'' formulation of quantum gravity \cite{perez:2013}.  In  Ref.~\cite{Dittrich_2016}, a tensor network formulation, as well as a coarse-graining scheme were developed for a spin-foam partition function.  In  Ref.~\cite{Asaduzzaman:2019mtx} the authors form a tensor network for the partition function of two-dimensional gravity where the gauge symmetry has been extended to merge the tetrad and spin-connection variables into a single connection.  In this work the Fisher zeros of the partition function are plotted in the complex-coupling plane.  
In a related work the authors in Ref.~\cite{bao2017} model de Sitter space using the multi-scale entanglement renormalization ansatz tensor network.
\item {\bf 
Other approaches} \\
As explained before, there is a large literature 
where MPS and PEPS are used to handle gauge theories \cite{Banuls:2019bmf,Banuls:2019rao} that is not reviewed here. For recents progress, see for instance \cite{emonts,funcke2020,frias2021,williamson2020}.
\end{itemize}

\section{Technologies and resources needed}
\begin{itemize}
\item
{\bf Method development} \\
A large variety of tensor network algorithms have been developed to 
approximately evaluate observables in Hamiltonian and Lagrangian systems 
(see section \ref{appendix:tm} in the appendix for a summary and references).
The methods span a range of compromises between accuracy, compute time and memory
use, with the most accurate methods generally requiring the most time and memory.
For systems with one spatial dimension, the time and memory requirements of the most accurate
methods can typically be accommodated on modern computers.
However, for three spatial dimensions, the computing demands for the most accurate methods
are well beyond what is presently possible, and a significant reduction in the
size and connectivity of the tensors within the network is necessary.
Several advances have been made to improve the scalability of methods,
which has made initial simulations of some 3+1 dimensional models possible.
However, high precision simulations of simple 3+1 dimensional systems
and initial tests for more complex theories (such as QCD) still remain a significant
challenge.
Sustained research into understanding the tradeoffs among approaches,
and developing new methods is necessary in order to
reach the accuracy and scalability needed for many of the problems of interest to HEP.

\item
{\bf Tensor network software and library requirements} \\
Tensor network calculations rely on efficient tensor (or matrix) math libraries.
The most common operations needed are
\begin{itemize}
    \item Reshaping the tensor and permuting the order of indices
    \item Element-wise operations (e.g. scaling elements, adding tensors)
    \item Contracting indices in a product of two (or more) tensors
    \item Tensor decompositions (typically SVD or eigen-decomposition)
\end{itemize}
A list of some tensor software can be found on the web \cite{tensornetwork.org}
and a more comprehensive list of packages is available here \cite{DBLP:journals/corr/abs-2103-13756}.
Many of the tensor libraries were developed and are currently used by the condensed matter community,
and may have MPS or other methods implemented.
Popular programming languages are C++, Python and, more recently, Julia.
For HEP research Numpy is still a popular choice for implementations of TRG and related algorithms.
Since many of the tensor libraries use optimized GEMM routines for contractions, performance
among them for large tensor sizes will be similar for the same set of operations.
The main differences between them tend to be in the interface (e.g. whether contractions use Einstein
notation, or some other helper functions).
This can make a difference in the ease of use, however, the order of contractions can make
a big difference in the memory and compute requirements, so interfaces that allow
specifying a collection of contractions need to either determine the optimal order or let the
user specify.
Determining the optimal contraction order becomes particularly important for larger networks
but is considered to be an NP-hard problem.
Finding efficient methods to determine nearly-optimal orderings 
is an active area of research, especially for classical simulations of quantum circuits.

{\bf Tensor decompositions}
In many cases the exact contraction of tensor networks is not feasible, and approximations must be made.
This is often done by using an SVD to reduce the size of intermediate tensors by
keeping only the largest modes, and hence having an efficient SVD is important.
Since typically only a small portion of the modes are being kept, a full SVD isn't necessary.
A partial SVD, such as one using randomized linear algebra \cite{halko2011finding}, can
be more efficient as the tensor sizes grow.

{\bf Distributed memory}
As the tensor sizes grow, the memory available on a single computer will become a bottleneck.
High accuracy calculations of theories such as QCD will likely require larger amounts of memory
found on leadership computing resources.
This will require having distributed memory tensor libraries that can efficiently perform the
necessary contractions and SVD procedures on distributed tensors.

{\bf Symmetries}
Many problems of interest to HEP have symmetries that can also be captured in the TN formulation.
This leads to sparse structure in the tensors which can be taken advantage of to make the
calculations more efficient and even more accurate.
Some tensor libraries developed for condensed matter applications can take symmetries into account.
Adopting and modifying these for problems in HEP could be beneficial.

{\bf Software ecosystem}
As noted in \cite{DBLP:journals/corr/abs-2103-13756}, there are a large number of libraries available
with overlapping capabilities, each developed for a specific research group or application in mind.
Due to the diverse set of target problems, methods and implementations, consolidating on a single
or small number of libraries may be difficult.
For HEP, having a coordinated research plan to develop tensor network software tailored for HEP problems,
while leveraging libraries actively developed in the condensed matter and computer science communities
may be a viable strategy to avoid further fragmentation.

\item
{\bf Access to large scale classical computers}\\
Unlike traditional lattice field theory methods 
which have a cost in compute time and memory that
grows with some power of the system volume, tensor network methods tend to have more modest resource growth for increasing volume.
Many TN methods have a resource growth that is only logarithmic in the volume, and some
can even work directly in the infinite volume limit.
The major factors influencing the resource requirements for TN methods are
\begin{itemize}
    \item The number of spatial dimensions in the theory
    \item The complexity of the theory (requiring a larger tensor on each site)
    \item The required accuracy of the final result
\end{itemize}
Increasing the accuracy of simulations requires growing the size of the intermediate tensors in the corresponding contraction and truncation procedure, to capture
the necessary entanglement in the system.

Improved truncation schemes can help reduce the memory footprint in favor of more compute cycles
(which can be advantageous as the compute power of computing technology tends to grow faster than memory
capacity).
However, when moving to 3+1 dimensions, the memory and compute requirements will both grow very fast
as the accuracy is improved, and high accuracy simulations of complex systems such as QCD will likely
require running across large scale leadership computing facilities.
In this case having efficient distributed memory versions of the codes will be necessary.
Having access to large machines (with large memory) will also be necessary to develop methods
and to calculate results.
Since there is no statistical sampling in TN methods, as opposed to standard lattice Monte Carlo methods, the total compute time needed could be less than for traditional lattice methods.
The main bottleneck, however, is likely to be the memory size needed to store tensors required for
the target accuracy.

In some cases, storage of the large tensors for later use may also be desired, requiring
sufficient storage space and I/O bandwidth for tensors of order of the size of the total
available memory.

\item
{\bf Workforce} \\
Progress on these problems will require a collaborative workforce across HEP and ASCR domains,
along with collaborations with other science domains such as condensed matter physics.
This will require skilled domain scientists familiar with the construction of lattice theories and relating numerical measurements to physical results.
It will also require computational scientists developing improved tensor network methods
that optimize the tradeoff between memory and compute time for a given
science problem and computer hardware.
They will also need to engage computer scientists and applied mathematicians who are developing
libraries and methods for large scale parallelization of tensor network contractions and decompositions.

\end{itemize}

\section{Overlap with other science domains and industry}
\begin{itemize}
\item
{\bf Condensed matter physics}\\
The development and application of tensor network methods is an active and vibrant
field in condensed matter physics.
More details on the background in condensed matter physics can be found in previous
sections and related references.
Many applications to HEP grew out of the related work by, and in collaboration with,
condensed matter researchers.
For example, the reformulation of lattice gauge theories was developed as part of a
collaboration extending the methods of Tao Xiang's group \cite{2012PhRvB..86d5139X}.
\item
{\bf Classical simulation of quantum circuits}\\
Tensor network methods can also be used to perform classical simulations of quantum circuits,
used for testing and development of quantum algorithms.
This can be done by using the tensor network to hold a compact approximation of the 
full state vector, for example using a PEPS network \cite{9355283}.
One can also use tensor network methods to simulate the output of quantum computers
without storing the full state vector \cite{Villalonga_2020,pan2021simulating}.
The classical simulation of quantum circuits is an important means of developing and
testing quantum algorithms for HEP applications.
Additionally, improvements in methods for simulating quantum circuits could have
an impact in tensor network methods directly applied to computing HEP problems,
and vice versa.
\end{itemize}

\section{Conclusions}

In summary, tensor network methods are playing an increasingly important 
role in several branches of physics and 
are also becoming important in the context of HEP. 
Tensor networks provide a compact representation of entangled quantum systems 
and quantum states that can be used to 
reformulate physical theories on a lattice, 
map them onto other quantum systems, such as quantum computers or quantum simulators,
and to classically evaluate simulation results either in the tensor network formulation,
or in the mapped representation onto quantum hardware.

Tensor Lattice Field Theory (TLFT) is a 
new approach to models studied in the context of lattice QCD. 
It provides reformulations of these models where the microscopic tensors are 
the translationally invariant, local, {\it building blocks} of 
exact discretizations of the path integral.
They encode the local and global symmetries of the original model.
They also 
provide a more general space of theories than the traditional Lagrangian and Hamiltonian densities.
Most model studied by lattice gauge theorists can be reformulated in TLFT.
Truncations need to be used for practical purposes and preserve symmetries in a generic way. This provides a broader class of models with the same universal continuum limit as the original model. 
TLFT methods connect smoothly between
the Lagrangian and Hamiltonian approaches and can be very useful in the 
context of quantum computing.

TLFT presents an increasingly attractive alternative to traditional
Monte Carlo methods, which are currently the only computational methods able to
provide first-principle calculations for a large range of HEP problems but
have limitations. In numerical methods based on coarse-graining, 
tensor network methods do not rely on statistical sampling, so can also provide solutions
to problems plagued by \emph{sign problems} that render Monte Carlo methods ineffective.
In some fields, such as condensed matter physics, their use is fairly common.
However, for problems of interest to HEP, 
tensor network methods are still very resource intensive, and are not currently feasible
for complex theories in four space-time dimensions, such as QCD.
Further development in tensor network methods and software is needed for them to make
a significant impact in HEP too.

Tensor networks can also provide a convenient representation for mapping
lattice theories to quantum hardware (either in terms of digital quantum circuits or
analog quantum simulators).
They can also be used to efficiently perform classical simulations of quantum circuits.
This is useful for developing and testing quantum computing algorithms, and is an important
area of research for determining the tipping point for quantum computers to provide a
quantum advantage over classical ones.
They help isolate ``building blocks'' that have a limited complexity and can be optimally approximated for NISQ hardware. 

We have summarized some of the key problems in HEP that could be addressed with tensor network methods, the current status of applications to HEP, and the developments in methods, software and personnel needed to make progress
in these areas.

Using TLFT for models directly relevant for HEP is a feasible goal for the next decade. 
Developing software and running codes in TLFT is labor intensive and requires
access to HPC. 


%

\appendix

\section{\label{appendix:tm}Brief background on tensor methods}

This is a brief overview of some tensor network methods.
More complete and detailed summaries can be found in several review articles
\cite{Orus_2019,Banuls:2019bmf,Banuls:2019rao,rmp,RevModPhys.93.045003}.

The Density Matrix Renormalization Group (DMRG) \cite{PhysRevLett.69.2863}
method (and the similar representation known as 
Matrix Product States (MPS) \cite{PhysRevB.55.2164})
was developed for 1D lattice Hamiltonians, and is now the most
accurate method available for many of these types of problems.
MPS can also be applied to 2D Hamiltonians, but this is typically limited
to cases with low entanglement.
The Projected Entanglement Pair States (PEPS) \cite{Verstraete2004,perez2007,schuch2010}
is an extension of MPS that can capture 2D entanglement well, but
at a greatly increased cost.

An alternative to MPS is provided by Tree Tensor Networks (TTN).
These use a binary tree of tensors to represent the state, which can be more efficient
in memory, especially for large systems, but generally does not provide as good a
representation as MPS.
This can be circumvented by the addition of \emph{disentanglers} into the tree network,
which provide extra connections within TTN that allow it to express much greater entanglement
(by applying unitary rotations that remove entanglement without changing the physics).
This is the basis of the Multiscale Entanglement Renormalization Ansatz (MERA) network \cite{PhysRevLett.99.220405}.
The disentanglers require extra effort to optimize, but for Hamiltonian systems, they still
allow for an efficient contraction of the network.

TTNs can be applied to 2D Hamiltonians as well, with and without disentanglers.
An implementation with disentanglers has been shown to work expecially well
\cite{PhysRevLett.126.170603}, however it comes with a large cost.

For HEP applications, an important problem is the evaluation of partition functions,
and related observables, that would come from Lagrangian systems (and also classical Hamiltonians).
One of the original methods designed for the evaluation of 2D partition functions
was the Tensor Renormalization Group (TRG) algorithm \cite{PhysRevLett.99.120601}.
This method explicitly decomposed and contracted neighboring tensors together
to produce a new renormalized tensor encoding the interactions of several sites.
An improved method for performing the tensor block transformation, 
based on a higher order SVD for tensors, was introduced as the Higher Order TRG 
(HOTRG) method \cite{2012PhRvB..86d5139X}.
This method has a relatively simple form and can easily be extended to higher
dimensions, though the entanglement is limited due to the tree-like structure of the
projections.

Several methods have been introduced to improve on the accuracy of HOTRG.
The Tensor Network Renormalization (TNR) method \cite{PhysRevLett.115.180405}
includes extra disentangling steps, similar to those used in MERA.
TNR has only been applied in 2D systems so far, and generalizing to higher dimensions is
non-trivial.
Other methods that attempt to remove unnecessary short-range degrees of while preserving
the long-range ones are Loop-TNR \cite{PhysRevLett.118.110504}
and Gilt-TNR \cite{PhysRevB.97.045111}.
These are based the idea of removing degrees of freedom around a loop
and are easier to apply in higher dimensions.

One method developed explicitly to scale well in higher dimensions is
Anisotropic TRG \cite{Adachi:2019paf}.
This is based on an update step that is similar to the original TRG method,
which is relatively cheap, but not as accurate as some of the above methods
at a given bond dimension.
Another variation developed to make the calculations even cheaper in memory
and compute time is to decomposes the network so that it only contains
triad tensors (ones with three indices) \cite{Kadoh:2019kqk}.
This representation can be combined with other methods such as HOTRG.

Choosing the best method for a given problem can be difficult as it depends on the details
of the observables being measured, the required accuracy and even the details
of the implementations and the computer hardware being used.
For problems in 4D, memory used is a significant constraint.
Methods that use less memory can be run with a larger bond dimension, than those
that consume more memory.
In some cases the larger bond dimension can make up for the accuracy lost (at fixed bond dimension) due to breaking up the tensor, or using a simpler update method.
However this is not always the case, since many of the methods are based on local update
steps which do not necessarily converge on the correct answer in a consistent manner.
In some cases a more expensive, but more uniformly converging method might give better
accuracy for a fixed hardware budget (memory and/or compute time).
Searching for the best combination of efficiency and convergence for problems
of interest to HEP is an open question.
\end{document}